%% file: panda_modeling.tex
\documentclass[journal]{IEEEtran}

\usepackage{amssymb}
\usepackage[cmex10]{amsmath}
\usepackage{amsfonts}
\usepackage{lineno}
\modulolinenumbers[5]
\usepackage{url}
\usepackage{hyperref}
\usepackage[plain]{fancyref}
\usepackage[noadjust]{cite}
\usepackage{textgreek}
\usepackage{tikz}
\usetikzlibrary{arrows, arrows.meta, decorations.pathreplacing}
\usepackage{graphicx}
\usepackage{grffile}
\graphicspath{{./fig/}}
\DeclareGraphicsExtensions{.pdf,.jpeg,.jpg,.png}
\usepackage{xcolor}
\usepackage{setspace}
\newcommand{\isot}[2]{\textsuperscript{#2}#1}

\usepackage{soul}
\usepackage{etoolbox}
\makeatletter
\patchcmd{\SOUL@ulunderline}{\dimen@}{\SOUL@dimen}{}{}
\patchcmd{\SOUL@ulunderline}{\dimen@}{\SOUL@dimen}{}{}
\patchcmd{\SOUL@ulunderline}{\dimen@}{\SOUL@dimen}{}{}
\newdimen\SOUL@dimen
\makeatother

\usepackage{url}
\usepackage[plain]{fancyref}
\usepackage{hyperref}

\makeatletter
\IEEEtriggercmd{\reset@font\normalfont\fontsize{8pt}{8pt}\selectfont}
\makeatother
\IEEEtriggeratref{1}

\begin{document}
\IEEEpubid{\begin{minipage}{0.85\textwidth}\ \\[12pt]\\ \\ \\ \centering
    \textcopyright 2023 IEEE.  Personal use of this material is permitted. Permission from IEEE must be obtained for all other uses, in any current or future media, including reprinting/republishing this material for advertising or promotional purposes, creating new collective works, for resale or redistribution to servers or lists, or reuse of any copyrighted component of this work in other works.
  \end{minipage}}

\title{Background and Anomaly Learning Methods for Static Gamma-ray Detectors}

\author{M.\,S.~Bandstra,
        N.~Abgrall,
        R.\,J.~Cooper,
        D.~Hellfeld,
        T.\,H.\,Y.~Joshi,
        V.~Negut,
        B.\,J.~Quiter,
        M.~Salathe,
        R.~Sankaran,
        Y.~Kim,
        and S.~Shahkarami%
\thanks{M.S. Bandstra, N.Abgrall, R.J.~Cooper, D.~Hellfeld, T.H.Y.~Joshi, V.~Negut, B.J.~Quiter, and M.~Salathe are with the Nuclear Science Division at Lawrence Berkeley National Laboratory, Berkeley, CA 94720 USA, e-mail: msbandstra@lbl.gov.
R.~Sankaran, Y.~Kim, and S.~Shahkarami are with the Argonne National Laboratory, Lemont, IL 60439 USA.}%
\thanks{This work was performed under the auspices of the U.S. Department of Energy by Lawrence Berkeley National Laboratory (LBNL) under Contract DE-AC02-05CH11231. The project was funded by the U.S. Department of Energy, National Nuclear Security Administration, Office of Defense Nuclear Nonproliferation Research and Development.}}

\maketitle


\begin{abstract}
Static gamma-ray detector systems that are deployed outdoors for radiological monitoring purposes experience time- and spatially-varying natural backgrounds and encounters with man-made nuisance sources.
In order to be sensitive to illicit sources, such systems must be able to distinguish those sources from benign variations due to, e.g., weather and human activity.
In addition to fluctuations due to non-threats, each detector has its own response and energy resolution, so providing a large network of detectors with predetermined background and source templates can be an onerous task.
Instead, we propose that static detectors use simple physics-informed algorithms to automatically learn the background and nuisance source signatures, which can them be used to bootstrap and feed into more complex algorithms.
Specifically, we show that non-negative matrix factorization (NMF) can be used to distinguish static background from the effects of increased concentrations of radon progeny due to rainfall.
We also show that a simple process of using multiple gross count rate filters can be used in real time to classify or ``triage'' spectra according to whether they belong to static, rain, or anomalous categories for processing with other algorithms.
If a rain sensor is available, we propose a method to incorporate that signal as well.
Two clustering methods for anomalous spectra are proposed, one using Kullback-Leibler divergence and the other using regularized NMF, with the goal of finding clusters of similar spectral anomalies that can be used to build anomaly templates.
Finally we describe the issues involved in the implementation of some of these algorithms on deployed sensor nodes, including the need to monitor the background models for long-term drifting due to physical changes in the environment or changes in detector performance.
\end{abstract}

\IEEEpeerreviewmaketitle%



\section{Introduction}
Detecting radioactive sources and/or nuclear material outside of regulatory control is an important problem that has been studied for many years~\cite{kouzes_detecting_2005}.
To address the search problem within an urban or populated area, some have considered placing static arrays or networks of detectors to provide coverage over a larger area than could be covered by a single mobile detector~\cite{nemzek_distributed_2004, stephens_detection_2004, brennan_radioactive_2005, deb_radioactive_2011, vilim_integrated_2011, cooper_intelligent_2012, rao_performance_2012, deb_iterative_2013, rao_network_2015, cazalas_defending_2018}.
In addition, some approaches for persistent monitoring of a city-sized area have included the placement of large, static sensors along busy roadways and at intersections~\cite{hoteling_analysis_2021}.
However, even if each sensor is nominally identical to all of the others in the array, the local radiological environment and slight differences in the detector hardware require approaches that either enable detection algorithms to adapt to each deployed system's idiosyncrasies, as well as to the temporal variability caused by nature and human activity, or that ignore these effects and tolerate degraded performance.
Additionally, encounters with sources, such as medical and industrial isotopes, might present differently in the spectral domain than expected from simulations due to the presence of material shielding the source in amounts and configurations that differ from prior assumptions, which may affect one's ability to correctly detect and identify those sources.

The natural backgrounds encountered by static detector systems in the outdoors are numerous.
The persistent portion of the background consists of the primordial isotope \isot{K}{40}, and the \isot{U}{238} and \isot{Th}{232} decay chains present in the soil and building materials around the detector (``KUT'' backgrounds) as well as contributions from cosmic rays~\cite{sandness_accurate_2009}.
The variable portions of the background consist of the \isot{Rn}{222} (\isot{U}{238} series) progeny \isot{Pb}{214} and \isot{Bi}{214} suspended in the air, which can vary with weather conditions; and rainfall, which creates particularly large increases in background through the wet deposition of \isot{Pb}{214} and \isot{Bi}{214} on the ground~\cite{damon_natural_1954, livesay_rain-induced_2014}.
Radon decay progeny from \isot{Rn}{220} (\isot{Th}{232} series) and \isot{Rn}{219} (\isot{U}{235} series) may also be present but are at negligible amounts due to their short half-lives (55.6\,s and 3.96\,s, respectively, versus \isot{Rn}{222}'s 3.82\,days), which inhibit their ability to diffuse out of the ground~\cite{nndc_nudat}.
Human activity around the detector can introduce nuisance sources in the form of medical isotopes (e.g., \isot{Tc}{99m}, \isot{I}{131}, and positron annihilation photons from isotopes such as \isot{F}{18}) and industrial isotopes (e.g., \isot{Cs}{137} and \isot{Ir}{192}).
Human activity can also create temporary decreases in the background through obstructing a detector's view of radioactive ambient materials, e.g., by vehicles~\cite{stewart_understanding_2018}.
The goal of this work is to enable a static detector to learn, over time and with minimal pre-set knowledge, the temporal and spectral patterns of its background and nuisance sources so that it can better distinguish threats.
Approaches are presented that use spectral and temporal features to learn about the background, and optionally with complementary data provided by a rain sensor.

This work was performed in the context of the Platforms and Algorithms for Networked Detection and Analysis (PANDA) project led by Lawrence Berkeley National Laboratory, which, in collaboration with the Domain Aware Waggle Network (DAWN) project led by Argonne National Laboratory, is developing and fielding a network of statically-deployed multi-sensor systems for radiological source detection, localization, and tracking in urban environments~\cite{cooper_networked_2023}.
In contrast to previous distributed detection systems, the PANDA-DAWN (PANDAWN) network integrates contextual sensors and radiation detectors, and features powerful edge computing to enable the deployment of real-time algorithms which can adapt to changing conditions in order to optimize and/or maintain performance.

In this work, methods are presented that allow detection algorithms for a single node in a network to adapt to the local radiological environment, including both spectral and temporal variations.
Since only single nodes are considered here, network effects are not explored.
Therefore these methods could be considered for any ``array'' of detectors (a series of nodes with minimal interactions between them), not just a true ``network'' (a series of detector nodes with collective interactions).
Multi-node network detection approaches are the subject of additional studies.


\section{Instrumentation}
\label{sec:instrumentation}
The instruments used for this work include NOVArray for the preliminary analysis, and PANDAWN nodes once they were deployed in the field.
Both systems will be described here.

NOVArray is an array of nineteen 2\(\times\)4\(\times\)16-inch NaI(Tl) detectors deployed at a height of approximately 2\,m on traffic light poles at major traffic intersections throughout Northern Virginia~\cite{hoteling_analysis_2021}.
NOVArray began collecting data in mid-2018.
The detectors collect 1024-bin spectra from 30\,keV to 3\,MeV at a rate of 1\,Hz and report these spectra along with status information for the digiBase photomultiplier tube (PMT) base, time, and GPS coordinates.
An overview of the array, as well as analysis of the typical nuisance sources observed, is presented in ref.~\cite{hoteling_analysis_2021}.

The next generation enhancement for systems like NOVArray is to include contextual data (e.g., visual imagery, weather sensors) to aid in the understanding of the radiological environment around each sensor in the array and the interpretation radiological anomalies, and also to enhance detection through networking with the other nodes.

\begin{figure*}
    \begin{center}
        \includegraphics[width=0.45\textwidth]{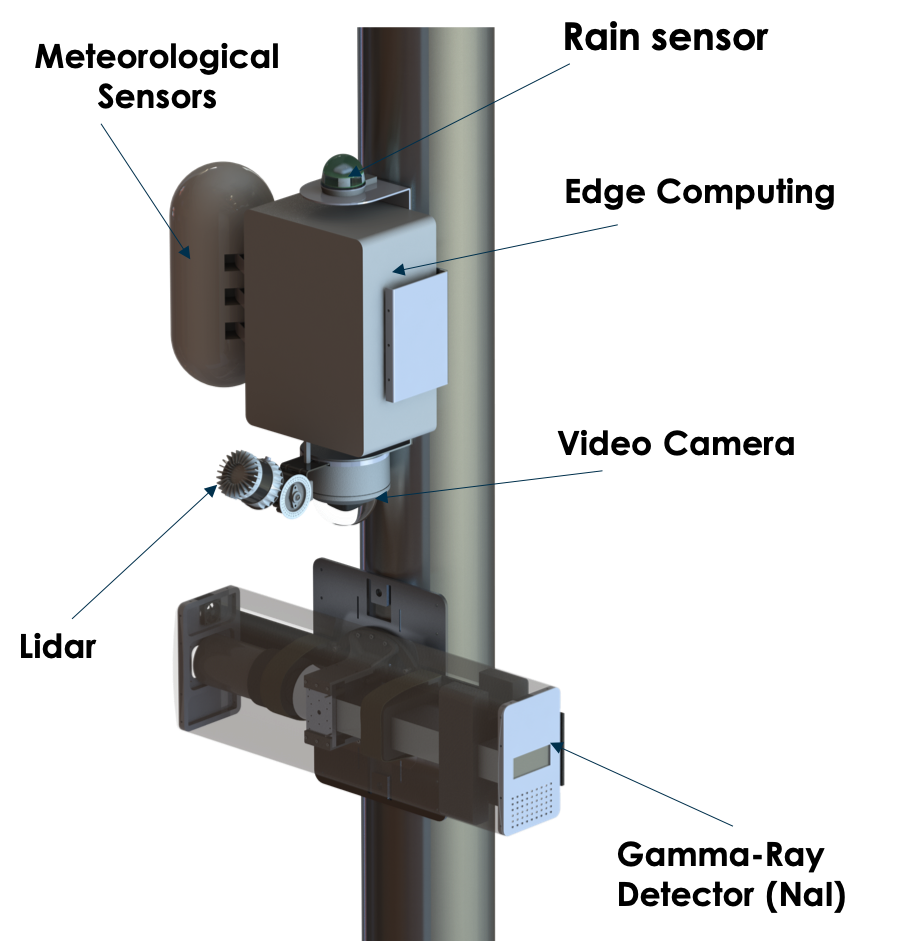}
        \includegraphics[width=0.33\textwidth]{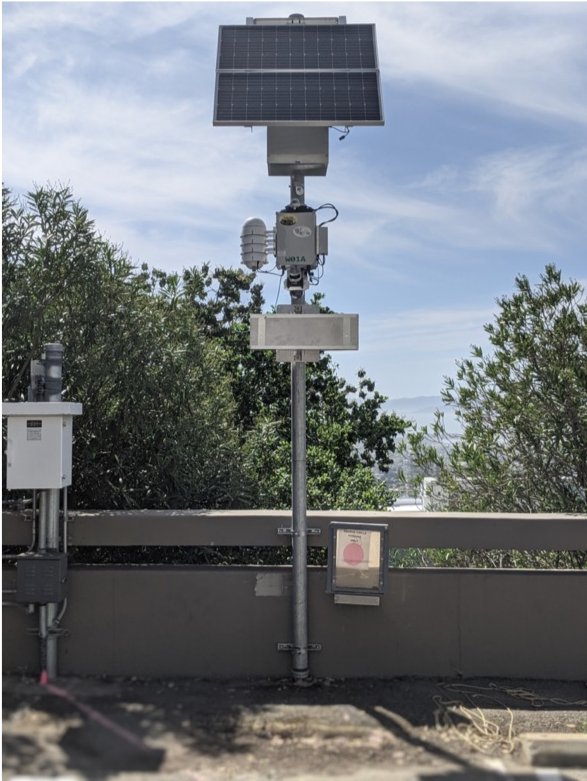}
    \end{center}
    \caption{A schematic of a PANDAWN detector node (left), shown with a prototype node mounted on a utility pole at LBNL (right). \label{fig:pandawn}}
\end{figure*}

The PANDAWN sensor systems consist of the same type of radiation sensors as NOVArray --- 2\(\times\)4\(\times\)16-inch NaI(Tl) detectors mounted on traffic poles and read out by digiBase PMT bases --- but are read out in list mode (i.e., as a time series of single event data) and are accompanied by additional contextual sensors.
These sensors include a light detection and ranging (LiDAR) unit in the form of an Ouster OS0 (Ultra-Wide model), a panoramic 5MP PTRZ camera (xnv-8081z model), a weather sensor (BME680 humidity, pressure and temperature sensor), and an optical rain sensor (Hydreon RG-15 model).
The sensors are read out by a software backend based on a customized job scheduler~\cite{sage_scheduler} within the Kubernetes framework.
Jobs handle data acquisition and on-the-edge processing which, at a lower level, are implemented as Robot Operating System (ROS) packages scheduled as Kubernetes Pods (referred to as ``plugins'').
The gamma-ray data are automatically calibrated in software by a novel algorithm that frequently refits the spectrum of recent gamma-ray events to a full-spectrum reference model and adjusts calibration parameters as necessary.
A schematic of a PANDAWN node is shown in~\Fref{fig:pandawn}, along with a photograph of a prototype node set up at LBNL for testing.
The first PANDAWN detector node built, tested, and deployed in Chicago was referred to as unit W022, and data from this system will be presented throughout this paper.
Other PANDAWN nodes are currently being deployed in Chicago, Illinois, USA\@.

The NOVArray data were used for exploratory research, and in this manuscript for analyses that require only spectral data and long timescales, since much less data were available for PANDAWN as of this writing.
In particular, data from NOVArray detector~17178694 will be shown, and its data were rebinned into 128 energy bins from 65 to 2920\,keV, with bin widths proportional to the square root of the energy to compress the data without losing energy resolution.
The endpoints of the spectra were chosen to avoid nonlinear calibration features at the low end, and the upper end was dictated by the gain and dynamic range of the readout of that specific detector.
Similarly, the PANDAWN data were binned from 30 to 3200\,keV in 200 bins with widths proportional to the square root of the energy.
The PANDAWN data used to demonstrate the methods were acquired from May 20, 2022 at 09:00 Central Standard Time (CST) to May 22, 2022 00:00 CST, for a total of 39\,hours.
The NOVArray data used were over a longer timespan, from between March 1, 2019 and September 30, 2019.

\section{Modeling static detector backgrounds}
In this section we will describe the framework we have chosen to analyze the data from static detectors, which consists of a combination of non-negative matrix factorization (NMF) in the spectral domain, and filtering in the time domain, optionally including rain sensor data.


\subsection{Non-negative matrix factorization}\label{sec:vanilla-nmf}
Non-negative matrix factorization (NMF)~\cite{lee_learning_1999, lee_algorithms_2001, paatero_positive_1994} was chosen as a framework for analyzing the spectral information from static gamma-ray detectors.
NMF is a dimensionality reduction method that decomposes a matrix of data (in this case, measured gamma-ray spectra) into a product of two non-negative matrices of (typically) much lower rank.
NMF is a natural choice for gamma-ray detector data because gamma-ray events are intrinsically positive quantities (i.e., photons do not constructively or destructively interfere with one another) and, neglecting detector effects like pileup, they are also linear (photons from different origins directly sum together in the measured spectrum).
In addition, unlike the more commonly known approach of principal component analysis (PCA), an NMF decomposition can be solved for by minimizing the Poisson likelihood between the data and the model, and thus NMF is compatible with the discrete nature of gamma-ray detector events.
NMF has already been successfully used as a framework for gamma-ray spectral anomaly detection and identification~\cite{bilton_nonnegative_2019}, as well as a method for understanding the different physical processes that give rise to background variability~\cite{bandstra_modeling_2020, bandstra_correlations_2021}.
The presentation of NMF here closely follows the presentation given in ref.~\cite{bandstra_correlations_2021}.

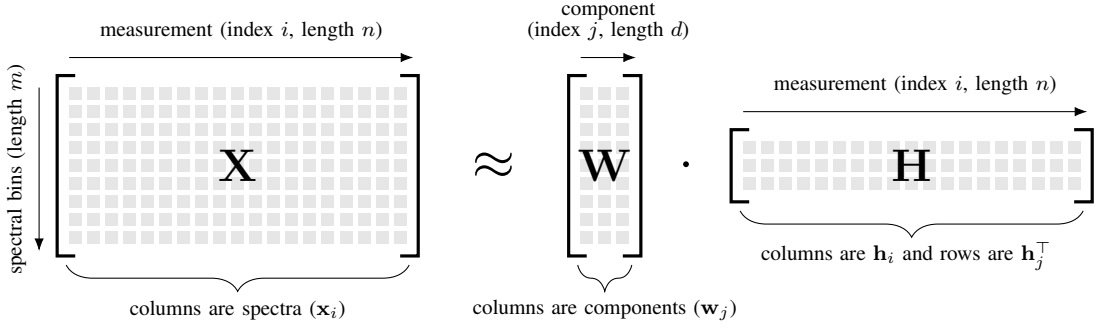
\begin{figure*}[t!]
    \begin{center}
        \input{fig/matrix_diagram.tex}
    \end{center}
    \caption{Diagram of the NMF decomposition (\fref{eq:nmf}) describing the matrix dimensions and the interpretation of the columns of each of the matrices. Diagram is from~\cite{bandstra_correlations_2021}. \label{fig:nmf}}
\end{figure*}

NMF decomposes an \(m \times n\) matrix of gamma-ray event counts (spectra) \(\mathbf{X}\), where \(m\) is the number of spectral bins and \(n\) is the number of spectra, into the product of two non-negative matrices of dimensions \(m \times d\) and \(d \times n\):
\begin{align}\label{eq:nmf}
    \mathbf{X}
        &\approx
            \mathbf{W}
            \mathbf{H}.
\end{align}
Here \(\mathbf{W}\) is the matrix of spectral components, \(\mathbf{H}\) is the matrix of time-dependent weights, and \(d\) is the number of NMF components.
These matrices are shown in~\Fref{fig:nmf}.

To make NMF compatible with Poisson statistics, \fref{eq:nmf} is solved by minimizing a cost function that is the negative log Poisson likelihood of the data given the model:
\begin{align}
    {\cal L}
        &=
            -
            \log L(
                \mathbf{X} | \mathbf{W}, \mathbf{H}
            )
        \equiv
            \sum\left(
                \mathbf{W}
                \mathbf{H}
                -
                \mathbf{X}
                \odot
                \log(
                    \mathbf{W}
                    \mathbf{H}
                )
            \right),
            \label{eq:nll}
\end{align}
where \( \odot \) is element-wise multiplication and the sum is taken over both matrix dimensions.
We have ignored terms that depend only on \(\mathbf{X}\) since they will not affect the solution.
Since in general our detectors operate at low dead time (\( \lesssim 1\% \)), we are also ignoring any instrumental effects that could lead to data that are overdispersed relative to the Poisson distribution.

The solution to~\fref{eq:nmf} that minimizes~\fref{eq:nll} can be arrived at using the maximum likelihood expectation maximization (MLEM) approach used by refs.~\cite{lee_learning_1999, lee_algorithms_2001}, which takes the form of the following multiplicative update rules:
\begin{align}
    \mathbf{W}
        &\leftarrow
            \mathbf{W}
            \odot
            \left(
                \frac{
                    \left(
                        \frac{
                            \mathbf{X}
                        }{
                            \mathbf{W}
                            \mathbf{H}
                        }
                    \right)
                    \cdot
                    \mathbf{H}^T
                }{
                    \mathbf{1}_{m, n} \cdot \mathbf{H}^T
                }
            \right)
            \label{eq:update_W}
        \\
    \mathbf{H}
        &\leftarrow
            \mathbf{H}
            \odot
            \left(
                \frac{
                    \mathbf{W}^T
                    \cdot
                    \left(
                        \frac{
                            \mathbf{X}
                        }{
                            \mathbf{W}
                            \mathbf{H}
                        }
                    \right)
                }{
                    \mathbf{W}^T
                    \cdot
                    \mathbf{1}_{m, n}
                }
            \right)
            \label{eq:update_H}
\end{align}
where \(\mathbf{1}_{m,n}\) denotes an \(m \times n\) matrix of ones.

Unlike approaches like PCA, with NMF one needs to specify \(d\) before training the model.
This issue can be remedied by training multiple models with different \(d\) values and selecting the best model using some criterion, such as the Akaike Information Criterion (AIC)~\cite{akaike_new_1974}, which has been used before for NMF model selection with radiation detector systems~\cite{bandstra_modeling_2020}.

A second disadvantage of NMF relative to PCA is that the solution to~\fref{eq:nmf} that minimizes~\fref{eq:nll} is typically not unique; any invertible \(d \times d\) matrix \(\mathbf{D}\) such that \(\mathbf{W}' \equiv \mathbf{W} \mathbf{D}^{-1}\) and \(\mathbf{H}' \equiv \mathbf{D} \mathbf{H}\) are both non-negative matrices means that the matrix product \(\mathbf{W}' \mathbf{H}'\) is an identical solution.
As a trivial example, any non-negative diagonal matrix would be a solution.
(PCA solutions are also technically not unique, since they can admit such transformations of their solutions, but \(\mathbf{D}\) can only be a diagonal matrix in that case.)
To eliminate that class of solution degeneracies, and to aid in the later interpretation of the model, at each step we will normalize the columns of \(\mathbf{W}\) so that they sum to unity.
This normalization can be accomplished by the matrix
\begin{align}
    \mathbf{D}
        &\equiv
            \mathrm{diag}(
                \mathbf{W}^{\top}
                \cdot
                \mathbf{1}_{m}
            )
            \label{eq:norm_1}
\end{align}
and the additional update rule
\begin{align}
    \mathbf{W}
        &\leftarrow
            \mathbf{W}
            \mathbf{D}^{-1}
            \label{eq:norm_3},
\end{align}
where \(\mathbf{1}_{m}\) is a length-\(m\) column vector of ones and \(\mathrm{diag}\) creates a diagonal matrix from a column vector.

A trivial example of an NMF model is a one-component model, which by definition results in the single column of \(\mathbf{W}\) converging to the mean spectrum normalized to sum to unity, and the single row of \(\mathbf{H}\) converging to the gross counts of each spectrum.
This kind of parsimonious model may often be relevant for static detectors when there is no rainfall since it can capture the persistent backgrounds caused by KUT sources and cosmic rays, and, in our experience, the variability of suspended \isot{Rn}{222} is not significant enough to require additional model complexity according to AIC\@.

When rainfall is present, a one-component model may often be disfavored by AIC in favor of more components.
Rainfall increases detector count rates due to rainout and washout of the \isot{Rn}{222} progeny \isot{Pb}{214} and \isot{Bi}{214} and results in spectral changes.
For this situation, two NMF components are typically required by AIC, as was the case for this dataset.
Models with more than one component were trained by randomly initializing the NMF components and applying the multiplicative update rules until the negative log likelihood changed by less than \(10^{-12} \times n\) or if more than 100,000 iterations had been performed --- the 2-component model converged in just over 15,000 iterations, while the 3-component model was stopped after 100,000, which took approximately 2,200 and 28,000\,s, respectively, using a 2.7\,GHz quad-core Intel i7.
With only random initializations, the temporal effect of rainfall in the two-component model was not completely isolated to the second component, but the two components contain a mixture of the persistent backgrounds and the rainfall background.
In principle, it would be better to have a cleaner separation between the two components to enable temporal modeling based on the dynamics of the rainfall contribution (i.e., addition of radioactivity from rainfall, decay of radioactivity according to the half-lives of \isot{Pb}{214} and \isot{Bi}{214}).
An approach to obtaining this outcome is presented in the following section.

\begin{figure*}
    \begin{center}
        \begin{tikzpicture}
            \draw (0, 0) node[inner sep=0] {\includegraphics[width=0.47\textwidth]{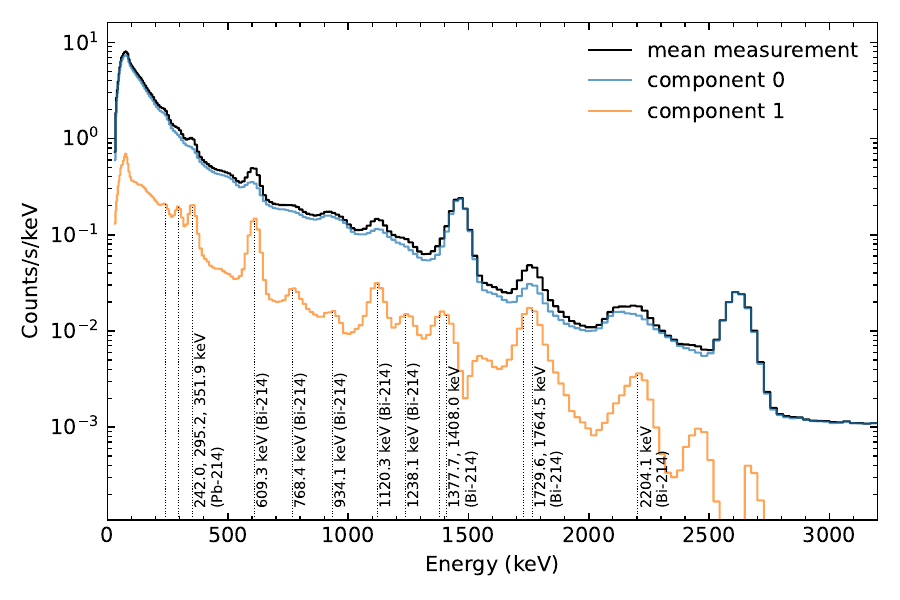}};
            \draw (-0.2, 2.0) node [text width=4.0cm, align=left] {\large Two components, \ \ \ \ \ \ \ \ regularized};
        \end{tikzpicture}
        \begin{tikzpicture}
            \draw (0, 0) node[inner sep=0] {\includegraphics[width=0.47\textwidth]{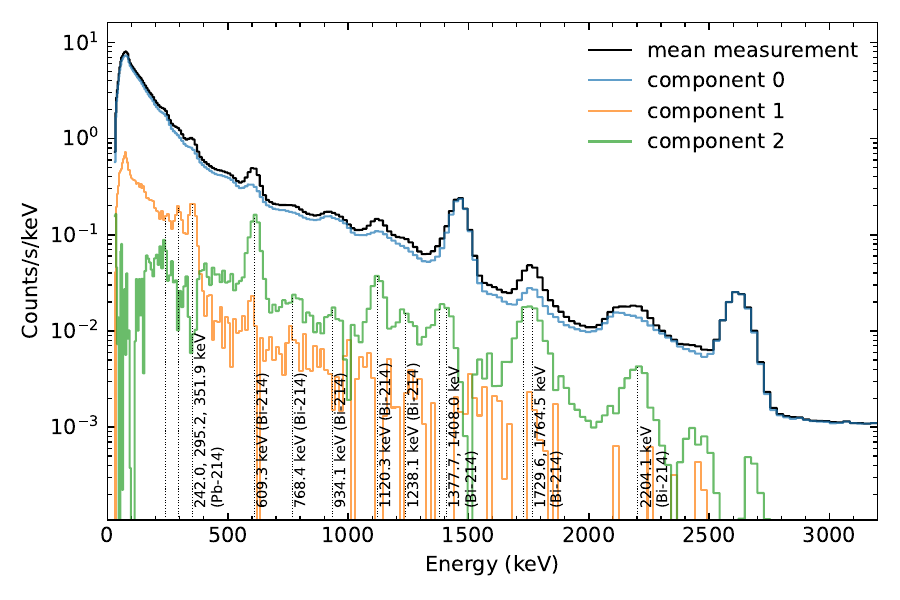}};
            \draw (-0.2, 2.0) node [text width=4.0cm, align=left] {\large Three components, \ \ \ \ \ \ regularized};
        \end{tikzpicture}\\
        \vspace{2mm}
        \begin{tikzpicture}
            \draw (0, 0) node[inner sep=0] {\includegraphics[width=0.96\textwidth]{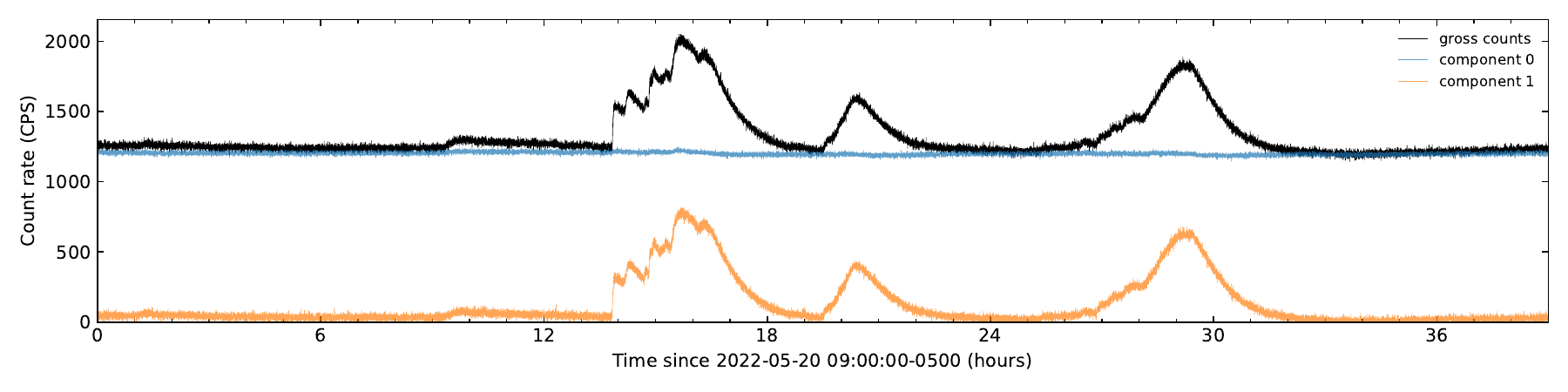}};
            \draw (-4.4, 1.3) node [text width=6.0cm, align=left] {\large Two components, regularized};
        \end{tikzpicture}\\
        \begin{tikzpicture}
            \draw (0, 0) node[inner sep=0] {\includegraphics[width=0.96\textwidth]{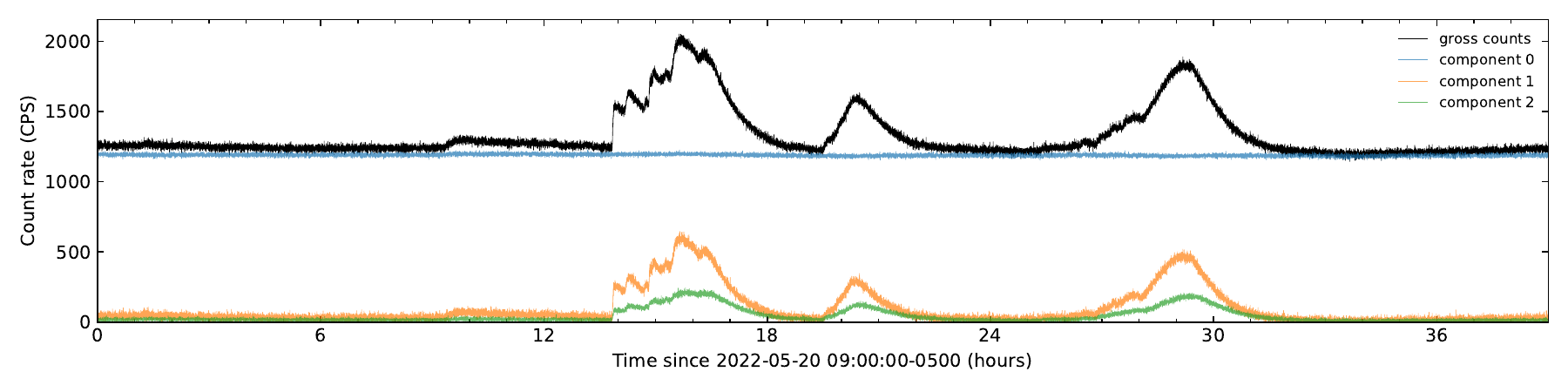}};
            \draw (-4.4, 1.3) node [text width=6.0cm, align=left] {\large Three components, regularized};
        \end{tikzpicture}
    \end{center}
    \caption{Different NMF models applied to data from the PANDAWN node W022 detector with a 5-second integration time.
    The single-component NMF model is not shown here but consists of a single spectral component proportional to the mean spectrum and a single weight component equal to the gross counts.
    The two-component model (spectral components at the top left, count rates versus time in the middle) is regularized to isolate the static background in component~0 and the rain-induced background in component~1.
    The three-component model (top right, and bottom) is regularized similarly and shows the separation between \isot{Pb}{214} and \isot{Bi}{214} temporal behavior during the rain events.
    In the top two plots of the model components, each component is weighted by the mean count rate derived from the corresponding row of the NMF weight matrix. \label{fig:nmf-models}}
\end{figure*}
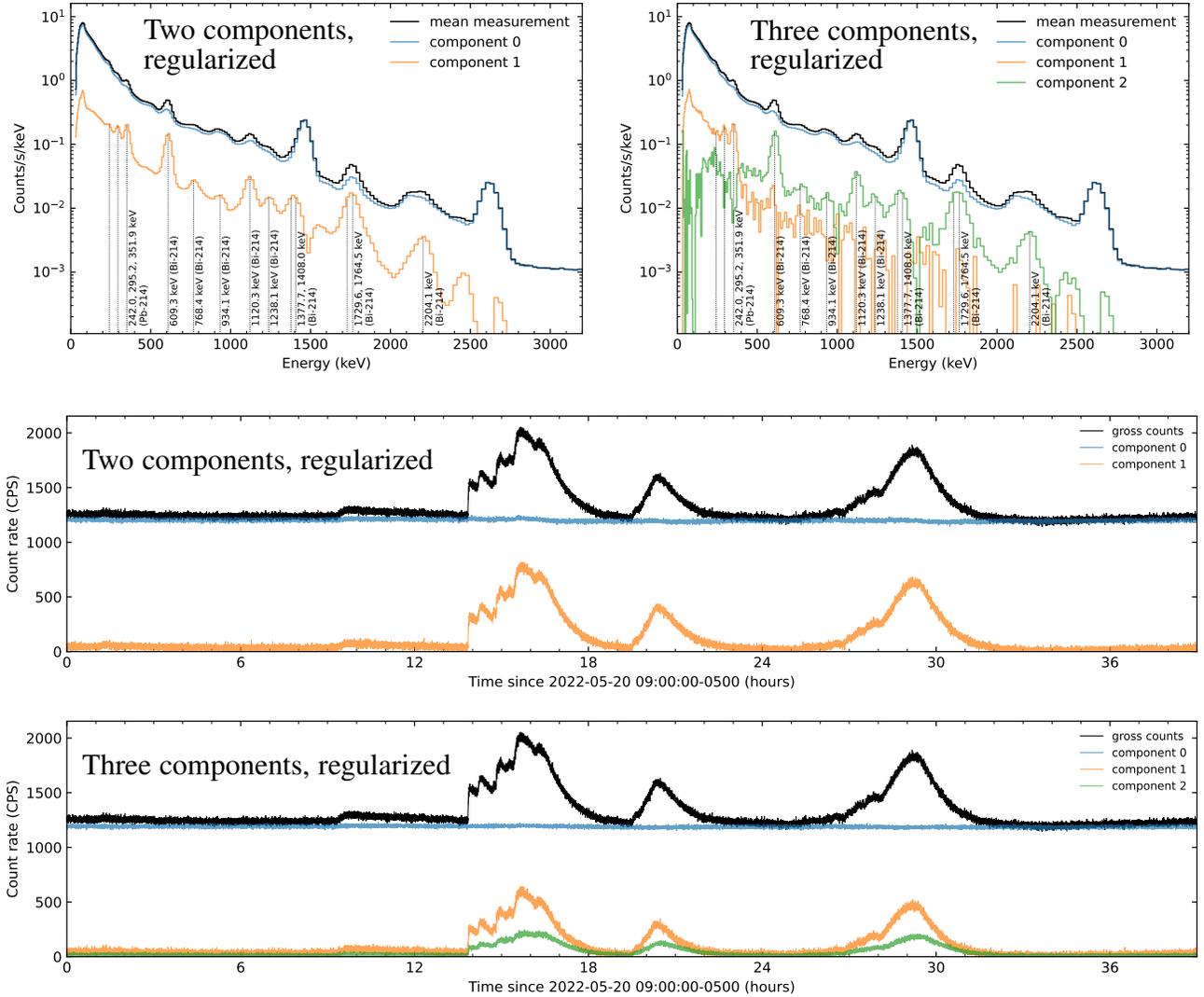


\subsection{Regularized NMF}\label{sec:regularized-nmf}
NMF models can be regularized during fitting so that the models have a desired property or properties.
An approach that has allowed the separation of rainfall-related (radon progeny) backgrounds from the persistent, static backgrounds is to apply regularization terms to the first NMF component's weights such that it gives that component a tendency to have a constant rate that is as large as possible, thus approximating the behavior of the static background.
To encode this property, we used two regularization functions to maximize the mean count rate of component~0 and minimize its variance:
\begin{align}
    f_1(\mathbf{H})
        &=
            -n \dot{\bar{H}}_0
            \label{eq:f1}
        \\
    f_2(\mathbf{H})
        &=
            \sum_{i=0}^{n-1}
                \left(
                    \frac{
                        H_{0i}
                    }{
                        \Delta t_i
                    }
                    -
                    \dot{\bar{H}}_0
                \right)^2,
\end{align}
where
\begin{align}
    \dot{\bar{H}}_0(\mathbf{H})
        &=
            \frac{1}{n}
            \sum_{i=0}^{n-1}
                \frac{
                    H_{0i}
                }{
                    \Delta t_i
                }
\end{align}
is the mean count rate for component 0 and \(\Delta t_i\) is the live time for spectrum \(i\).

The gradients of these regularization functions are
\begin{align}
    \left(
        \nabla_{\mathbf{H}}
        f_1
    \right)_{pq}
        &=
            -
            \frac{
                \delta_{p0}
            }{
                \Delta t_{q}
            }
        \equiv
            -
            \left(
                \nabla_{\mathbf{H}}
                f_1
            \right)_{pq}^{-}
        \\
    \left(
        \nabla_{\mathbf{H}}
        f_2
    \right)_{pq}
        &=
            2
            \frac{
                \delta_{p0}
            }{
                \Delta t_{q}
            }
            \left(
                \frac{
                    H_{pq}
                }{
                    \Delta t_{q}
                }
                -
                \dot{\bar{H}}_0
            \right)
        \equiv
            \left(
                \nabla_{\mathbf{H}}
                f_2
            \right)_{pq}^{+}
            -
            \left(
                \nabla_{\mathbf{H}}
                f_2
            \right)_{pq}^{-},
\end{align}
where \( \delta \) is the Kronecker delta and we have grouped the algebraic terms so that all of the positive algebraic terms are in the gradient denoted by \(+\) and all the negative terms are in the gradient denoted by \(-\), so thus each of the gradient terms is positive.

The multiplicative update rule for~\(\mathbf{H}\) can be modified to approximately minimize these two functions while also minimizing the Poisson loss (e.g.,~\cite{becker_adaptive_2015}).
To do this we adjust~\fref{eq:update_H} to be
\begin{align}
    \mathbf{H}
        &\leftarrow
            \mathbf{H}
            \odot
            \left(
                \frac{
                    \mathbf{W}^T
                    \cdot
                    \left(
                        \frac{
                            \mathbf{X}
                        }{
                            \mathbf{W}
                            \mathbf{H}
                        }
                    \right)
                    +
                    \alpha_2
                    \boldsymbol\nabla_{\mathbf{H}}^{-}
                    f_2(\mathbf{H})
                }{
                    \mathbf{W}^T
                    \cdot
                    \mathbf{1}_{m, n}
                    +
                    \alpha_1
                    \boldsymbol\nabla_{\mathbf{H}}^{+}
                    f_1(\mathbf{H})
                    +
                    \alpha_2
                    \boldsymbol\nabla_{\mathbf{H}}^{+}
                    f_2(\mathbf{H})
                }
            \right),
            \label{eq:update_H_reg}
\end{align}
where \(\alpha_1\) and \(\alpha_2\) are the weightings applied to each of the regularization terms.

A two-component model using these regularizations to isolate the radon progeny component is shown in~\Fref{fig:nmf-models} using PANDAWN data with a 5-second integration time.
The model was found after optimizing for just over 45,000 iterations of the multiplicative update rules, which took approximately 6800\,s with a 2.7\,GHz quad-core Intel i7.
After trying values over several different orders of magnitude, \(\alpha_1 = \alpha_2 = 10^{-3}\) were found to induce the desired effects in the NMF model without forcing the effects too strongly, e.g., causing the elements in the first row of \(\mathbf{H}\) to be exactly the same number.
It can be seen that of the two components, component~0 captures most of the background features, but component~1 captures those features specific to the radon progeny that are measured in excess of the static background during rain events: the 295 and 352\,keV lines of \isot{Pb}{214}, and the 609, 1120, 1764, and 2204\,keV lines of \isot{Bi}{214}, among several others.
For this model, the elements of \(\mathbf{W}\) and \(\mathbf{H}\) were randomly initialized in the interval \( [ 10^{-4}, 1 ] \), and so the emergence of a component~1 that isolates the radon progeny is only an effect of the regularization functions and the physics of the situation.

In addition, a three-component model was also trained with the same regularization and random initialization.
This model was stopped after 100,000 iterations, which took approximately 15,300\,s with a 2.7\,GHz quad-core Intel i7.
As seen in~\Fref{fig:nmf-models}, the excess activity during the rain is now shared by components~1 and 2, and the component spectra reveal a separation between the \isot{Pb}{214} lines, which are only present in component~1, and the \isot{Bi}{214} lines, which are only present in component~2.
This separation was not included in the regularizations (which merely maximized and held component~0 nearly constant) but reflects the actual evolving ratio between the two isotopes as the rain falls and the two isotopes decay on the ground (\isot{Pb}{214} decays with a half-life of 27.06\,min directly into \isot{Bi}{214}, which has a half-life of 19.9\,min)~\cite{nndc_nudat}.
The ratio of \isot{Bi}{214} to \isot{Pb}{214} activity is expected to evolve substantially over time in the precipitation itself and in the accumulated rainwater on the ground.
The smallest ratio of their activities is 1.0, which they reach in secular equilibrium in the cloud droplets; it then evolves over the tens of minutes the raindrops take to fall, and after multiple hours the ratio will reach its largest (transient equilibrium) value of 3.88~\cite{greenfield_determination_2008}.


\subsection{Start-up procedure: spectral ``triage''}
\label{sec:spectral-triage}
The previous sections describe spectral analysis that one could retrospectively perform on collected data to determine a spectral model for the data.
However, in a fielded system, it is desirable to develop a model during data collection, and perhaps even update the model as more data are collected.

To approach this task, we decided to exploit temporal information that is complementary to the spectral information described in the previous sections.
The simple method we developed and tested is referred to as ``spectral triage,'' and uses slow and fast exponentially weighted moving average (EWMA) gross count-rate filters, and optionally a rain sensor if available, to decide which of the three categories an incoming spectrum is likely to be a member of:
\begin{enumerate}
    \item Static background (no triggers)
    \item Rain event (slow gross count-rate trigger only)
    \item Nuisance or threat (both alarms are triggered)
\end{enumerate}
Then the spectra from each of these three groups can be used for further analysis and model construction.
In this work, to construct spectral models, we sum the spectra in group~1 and normalize by their total integration time to find the static background rate; and we sum the spectra in group~2, normalize by time, and then subtract the static background rate to obtain the rain component rate.
We also demonstrate in~\Fref{sec:anomaly-clustering} how clustering techniques could be applied to the spectra in group~3 to classify the types of anomalies encountered.
Note that one could additionally apply a spectral anomaly detection method such as N-SCRAD or Censored Energy Window as an extra check for anomalous spectra~\cite{detwiler_spectral_2015, pfund_improvements_2016, lei_robust_2017, miller_gammaray_2018} that would aid in the categorization specified above.

The timescales used for the slow and fast EWMA filters in spectral triage are 120~minutes and 30~seconds, respectively.
(These timescales are related to the filter's exponential parameter \(\lambda\) by \(\lambda = \Delta t / T\), where \(\Delta t\) is the integration time of the sensor and \(T\) is the timescale.)
The slow timescale is chosen to allow for any diurnal variations in background to be followed.
Such variations may include radon variations from weather or slight changes in the registered gross count rate due to temperature-induced gain shifts in the detector readout electronics.
These gain shifts result in changes in the number of events exceeding the low-energy threshold which then manifest as gross count shifts.
The shorter timescale is chosen to allow the fast filter to follow any increases in gross count rate due to rainfall, and could eventually be informed by an attached rain sensor if available.


\subsubsection{EWMA filters}\label{sec:ewma}

For spectral triage, each EWMA filter is calculated to track the mean gross count rate \(\mu\) and the variance of the measured count rate \(\sigma^2\) according to some time constant \(\lambda\).
If the gross counts of measurement \(k\) is \(n_k\), with livetime \(\Delta t_k\), the filter is updated as follows.
For the first iteration, because of the Poisson nature of the measurement,
\begin{align}
    \hat{\mu}_0
        &=
            \frac{
                n_0
            }{
                \Delta t_0
            },
        \\
    \hat{\sigma}^2_0
        &=
            \frac{
                n_0
            }{
                \Delta t_0^2
            }.
\end{align}
Then on subsequent iterations starting at \(k=1\), first the following z-score (approximately unit normal) statistic is formed to check for anomalous count rates:
\begin{align}
    z_k
        &=
            \frac{
                n_k
                -
                \hat{\mu}_{k-1}
                \Delta t_k
            }{
                \sqrt{
                    \max
                    \left(
                        \hat{\sigma}^2_{k-1}
                        \Delta t_k^2,
                        \hat{\mu}_{k-1}
                        \Delta t_k
                    \right)
                }
            }.
\end{align}
The denominator contains the maximum of two terms since \( \hat{\sigma}^2_{k-1} \Delta t_k^2 \) on average should estimate the variance of the numerator, but it is clipped at the minimum value expected from Poisson statistics (\( \hat{\mu}_{k-1} \Delta t_k \)).
This same type of clipping is used when tracking the variance of the mean spectrum by the N-SCRAD algorithm~\cite{pfund_improvements_2016}.

Before updating the filter for the \(k\) step, the \(k\)-th data point is checked to see whether it is anomalous.
There are two kinds of anomalies.
If \( | z_k | > z_{\mathrm{filter}} \), then the \(k\)-th data point is not used to update the filter so as not to contaminate the filter.
The second type of anomaly is when \(z_k > z_{\mathrm{alarm}}\), and then the filter records an ``alarm'' (and is also not updated).
In general \(z_{\mathrm{filter}} \le z_{\mathrm{alarm}}\), and we chose values of \(z_{\mathrm{filter}}\) of 2 or 3 depending on how conservative the filter should be when incorporating new data.
The value of \(z_{\mathrm{alarm}}\) can be set to the desired false positive probability or false alarm rate by assuming it has a unit normal distribution.
Positive and negative deviations are considered for ``filter'' anomalies since positive deviations may be the result of sub-alarm-threshold sources, and negative deviations can frequently occur as the result of the movement of vehicles and people nearby.
Both types of changes should be prevented from contaminating the filter.
For ``alarm'' anomalies, only positive deviations are considered since only those anomalies could be the result of source emission.

If the count rate is not anomalous in either of these two ways, then the EWMA update step is performed:
\begin{align}
    \hat{\mu}_k
        &=
            (1 - \lambda)
            \hat{\mu}_{k-1}
            +
            \lambda
            \frac{
                n_k
            }{
                \Delta t_k
            }
        \\
    \hat{\sigma}^2_k
        &=
            (1 - \lambda)
            \hat{\sigma}^2_{k-1}
            +
            \lambda
            \left(
                \frac{
                    n_k
                }{
                    \Delta t_k
                }
                -
                \hat{\mu}_{k-1}
            \right)^2.
\end{align}

This algorithm is a variation of the commonly used ``K-Sigma'' algorithm~\cite{jarman_comparison_2008, bilton_nonnegative_2019}.


\subsubsection{Spectral triage results}
\label{sec:triage-results}

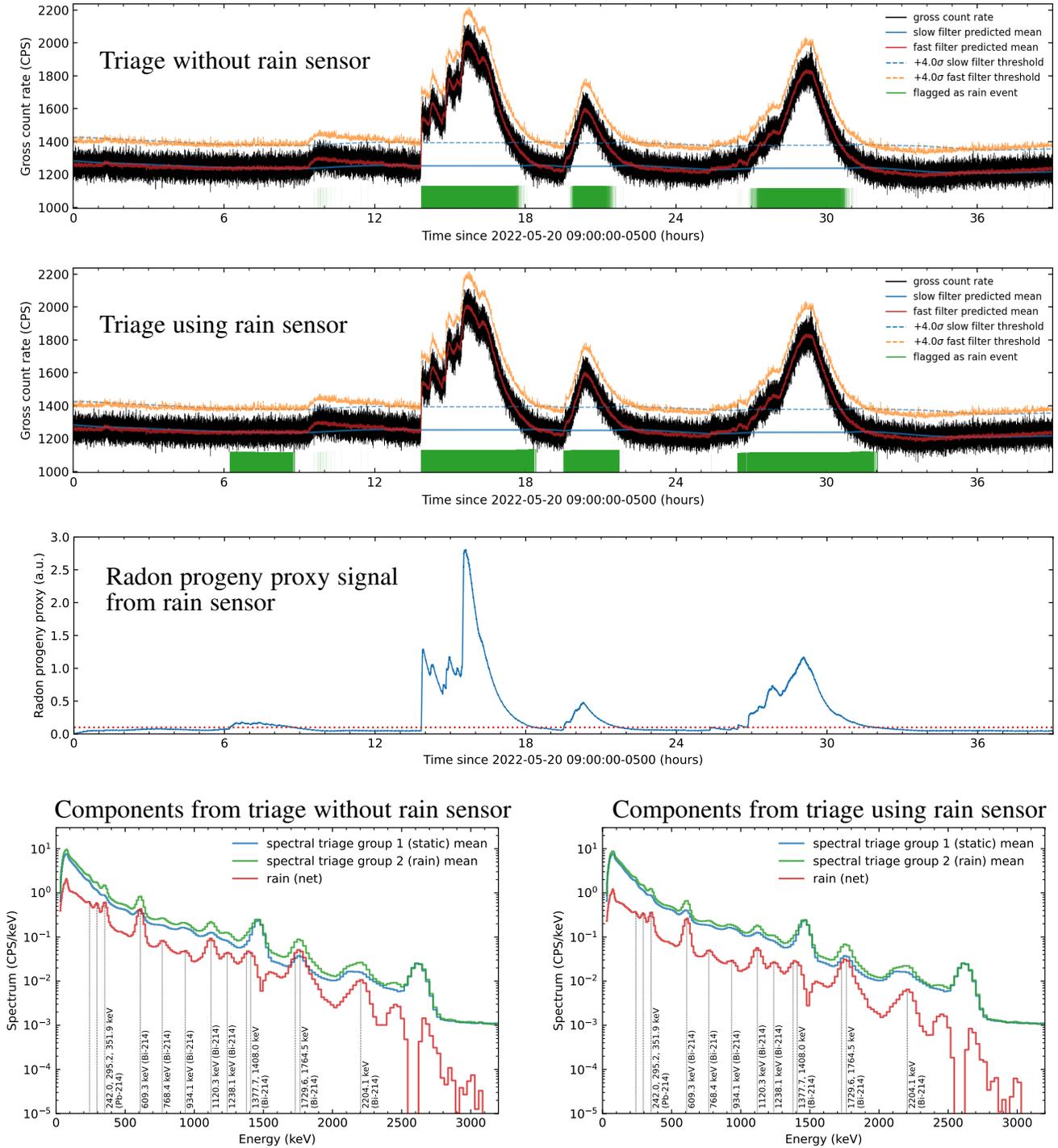
\begin{figure*}
    \begin{center}
        \begin{tikzpicture}
            \draw (0, 0) node[inner sep=0] {\includegraphics[width=0.96\textwidth]{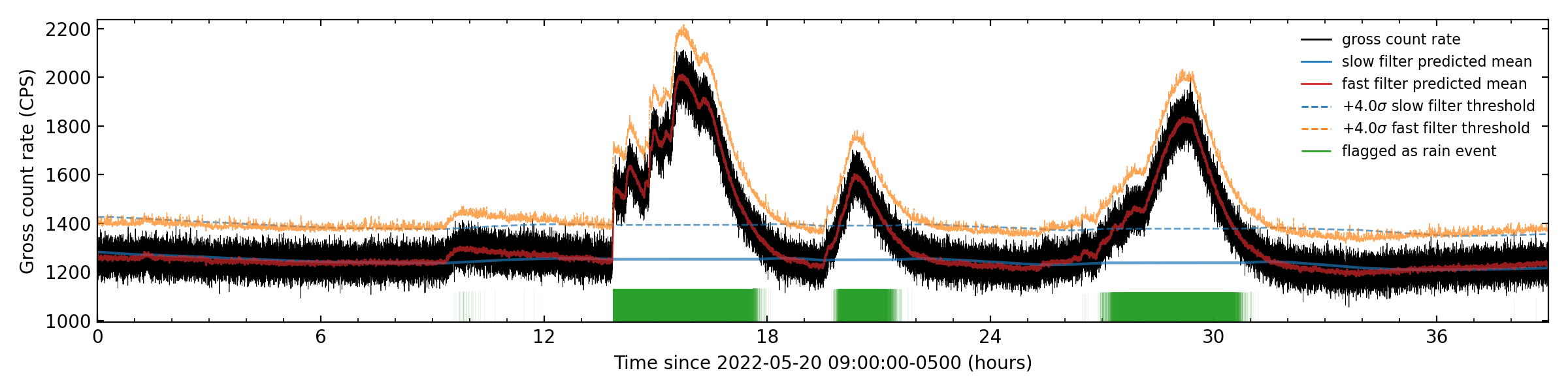}};
            \draw (-4.2, 1.0) node [text width=6.0cm, align=left] {\large Triage without rain sensor};
        \end{tikzpicture}\\
        \begin{tikzpicture}
            \draw (0, 0) node[inner sep=0] {\includegraphics[width=0.96\textwidth]{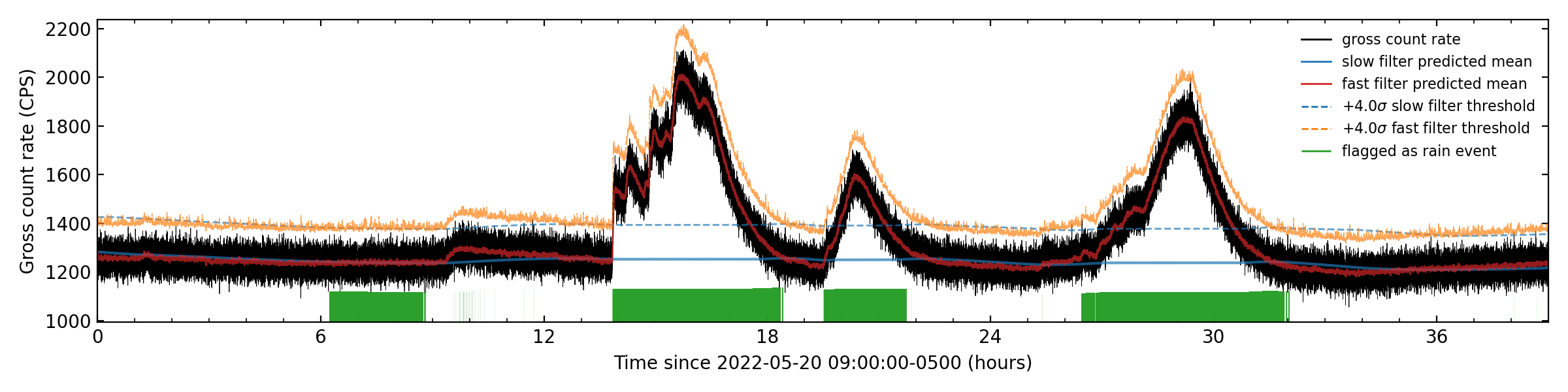}};
            \draw (-4.2, 1.0) node [text width=6.0cm, align=left] {\large Triage using rain sensor};
        \end{tikzpicture}\\
        \hspace{0.005\textwidth}
        \begin{tikzpicture}
            \draw (0, 0) node[inner sep=0] {\includegraphics[width=0.95\textwidth]{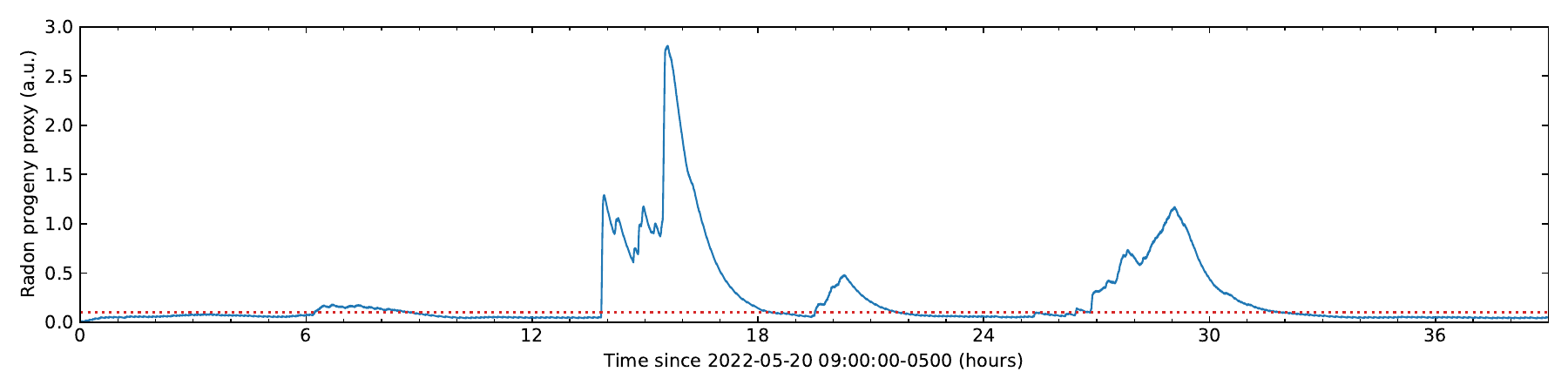}};
            \draw (-4.2, 1.0) node [text width=6.0cm, align=left] {\large Radon progeny proxy signal \ \ \ \ from rain sensor};
        \end{tikzpicture}\\
        \vspace{2mm}
        \begin{tikzpicture}
            \draw (0, 0) node[inner sep=0] {\includegraphics[width=0.47\textwidth]{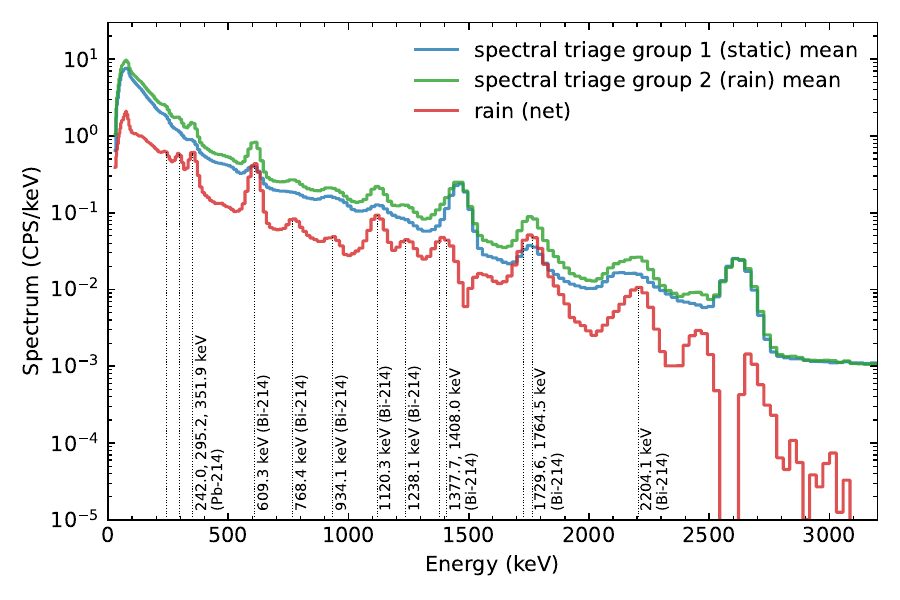}};
            \draw (0.5, 2.9) node [text width=8.0cm, align=center] {\large Components from triage without rain sensor};
        \end{tikzpicture}
        \begin{tikzpicture}
            \draw (0, 0) node[inner sep=0] {\includegraphics[width=0.47\textwidth]{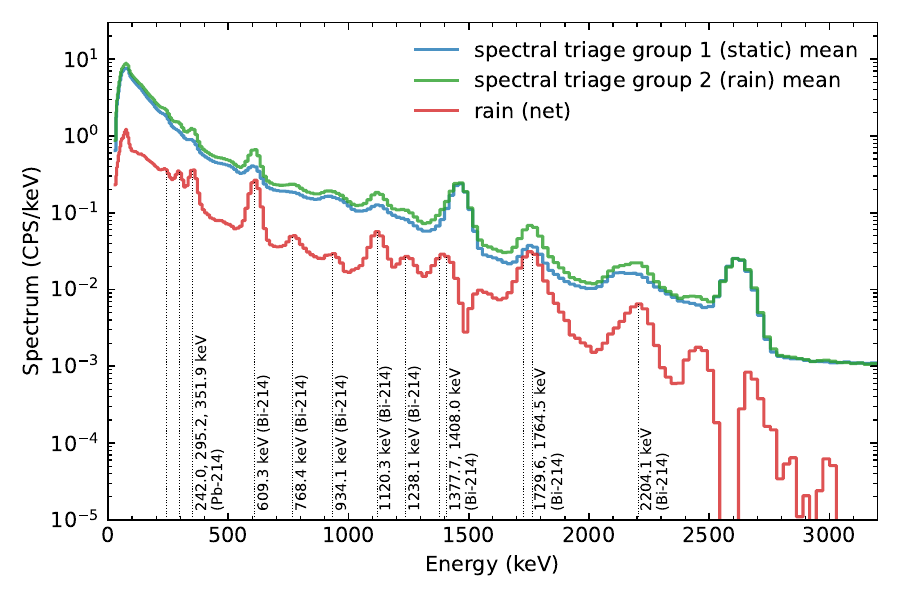}};
            \draw (0.5, 2.9) node [text width=8.0cm, align=center] {\large Components from triage using rain sensor};
        \end{tikzpicture}
    \end{center}
    \caption{Spectral triage results for PANDAWN node W022 detector.
    The top plot shows which spectra are labeled as rainfall-related (triage group~2) using the standard (not rain-aware) version of the algorithm, while the second plot from the top includes the rain-sensor logic, derived from the radon progeny proxy quantity shown in the third plot.
    At the bottom are the spectral models derived from the standard spectral triage (left) and rain-aware spectral triage (right), which result in nearly identical spectral decompositions.
    The integration time for the data was 1\,s. \label{fig:triage-results}}
\end{figure*}

An example of using spectral triage on the same data as~\Fref{fig:nmf-models} is shown in~\Fref{fig:triage-results}.
Here the integration time of the sensor was 1\,s, so the EWMA parameters were \(\lambda_{\mathrm{slow}} = 6.944 \times 10^{-4}\) and \(\lambda_{\mathrm{fast}} = 2.78 \times 10^{-2}\).
For the slow and fast filters, we chose \(z_{\mathrm{filter}}\) = 2 and 3, respectively.
Using a false alarm rate of 1 in 8~hours, and given an integration time of 1\,s, \(z_{\mathrm{alarm}}\) for both the slow and fast filters was calculated to be 3.978 assuming only positive deviations count toward the false alarm rate.

\subsubsection{Spectral triage using a rain sensor}
\label{sec:rain-sensor}

For the PANDAWN W022 node, the rain sensor recorded accumulated rain at a rate of 1\,Hz.
Since the radon progeny in rain take time to radioactively decay once they have fallen on the ground, the rain sensor itself is not a good indicator of when the count rates from rain will be elevated.
Instead, we formed a ``radon progeny proxy'' model that filters the rainfall rate data \( r_k \):
\begin{align}
    r^{\mathrm{proxy}}_0
        &=
            r_0
        \\
    r^{\mathrm{proxy}}_k
        &=
            r^{\mathrm{proxy}}_{k-1}
            e^{-\lambda_B (t_k - t_{k-1})}
            +
            r_k
        \\
    &\approx
        \left[
            1
            -
            \lambda_B
            (t_k - t_{k-1})
        \right]
        r^{\mathrm{proxy}}_{k-1}
        +
        r_k
\end{align}
where \(\lambda_B\) is the decay constant of \isot{Pb}{214} (a.k.a. Radium B), which is \(4.2692 \times 10^{-4}\)\,s\(^{-1}\).
This model assumes that the activity concentrations of both \isot{Pb}{214} and \isot{Bi}{214} in the falling raindrops is constant over time, and that their ratio is also constant, which are not good assumptions in general, as discussed earlier in Sections~\ref{sec:vanilla-nmf} and \ref{sec:regularized-nmf} and in references~\cite{finck_situ_1980, paatero_wet_2000, takeyasu_concentrations_2006, greenfield_determination_2008}.

When the radon progeny proxy quantity exceeds a threshold, we consider that to be a rain event, and this alarm is included with a logical OR with the existing logic for triage group~2.
The threshold for this detector was \(0.10\) and was set from examining its behavior over long periods without rain.
An example of the calculated radon progeny proxy quantity for PANDAWN node W022 data, and the corresponding spectra identified as rainfall-related, is shown in~\Fref{fig:triage-results}.
Note that the rain sensor detects rain around hours~6--9 in that dataset, but there is no corresponding increase in count rate.
However, there is an increase in count rate at around hour~10 that is detected by the standard algorithm but little rainfall is detected by the sensor.
The causes of these discrepancies is unknown, but for the latter it is possible that the pole that the system is mounted on could have shielded the rain sensor from the rain.
Because of this potential issue, we have decided to run spectral triage with both the standard logic and the rain sensor OR-ed together.


\section{Anomaly clustering}
\label{sec:anomaly-clustering}
The spectra in the triage group~3, which are possible nuisance or threat anomalies, can be sent to a clustering algorithm.
The desired output of an anomaly clustering algorithm is a set of representative spectra corresponding to each type of anomaly source encountered, with the goal being to find shapes that are more representative of the sources than could readily be inferred from, e.g., Monte Carlo simulations alone, due to factors such as encountering different attenuation and scattering by the vehicles on the road than those considered in simulations.
Here we will present two kinds of algorithms that were explored and discuss their advantages and disadvantages.


\subsection{K-means clustering with KL divergence}\label{sec:cluster-kldiv}
A relatively straightforward approach to clustering was implemented with some success.
All of the potentially anomalous spectra were normalized to sum to unity, and K-means clustering was used to cluster the spectra into different groups.
The Kullback-Leibler (KL) divergence~\cite{kullback_information_1951} was used as distance metric between each normalized spectrum (\(\mathbf{p}\)) and cluster center (\(\mathbf{q}\)):
\begin{align}
    D_{\mathrm{KL}}(
        \mathbf{p}
        \|
        \mathbf{q}
    )
        &=
            \sum_i
                p_i
                \log
                \left(
                    \frac{
                        p_i
                    }{
                        q_i
                    }
                \right).
\end{align}

The KL divergence is expected to be useful here because it is approximately the negative log Poisson likelihood divided by the gross counts.
Specifically, if the spectrum is \(\mathbf{x} = N \mathbf{p}\) where \(N = \sum_i x_i \) are the gross counts of the spectrum and \(\mathbf{p} = \mathbf{x} / N\), and the cluster center \(\mathbf{q}\) is scaled with some parameter \(\alpha\) to maximize the Poisson likelihood, the result is that \(\alpha = N\) maximizes the likelihood.
The resulting negative log likelihood is
\begin{align}
    -&\log L(
        N \mathbf{p}
        \|
        N \mathbf{q}
    ) \\
    &=
        \sum_i
            \left[
                N q_i
                -
                N p_i
                \log(
                    N q_i
                )
                +
                \log(
                    N p_i
                )!
            \right]
    \\
    &\approx
        \sum_i
            \left[
                N q_i
                -
                N p_i
                \log(
                    N q_i
                )
                +
                N p_i
                \log (
                    N p_i
                )
                -
                N p_i
            \right]
    \\
    &=
        N
        \sum_i
            p_i
            \log\left(
                \frac{
                    p_i
                }{
                    q_i
                }
            \right)
    \\
    &=
        N
        D_{\mathrm{KL}}(
            \mathbf{p}
            \|
            \mathbf{q}
        )
\end{align}

For K-means we need to be able to calculate the cluster center (or mean) from a cluster of normalized spectra.
To define the cluster center \(\mathbf{q}\) for a cluster of normalized spectra \(\{\mathbf{p}_j\}, j = 0..n-1\), we find the value of \(\mathbf{q}\) that minimizes the KL divergence to all of the normalized spectra.
In other words, we find the normalized vector \(\mathbf{q}\) that minimizes:
\begin{align}
    D_{\mathrm{cluster}}(
        \{\mathbf{p}_j\},
        \mathbf{q}
    )
        &\equiv
            \sum_{j=0}^{n-1}
                D_{\mathrm{KL}}(
                    \mathbf{p}_j
                    \|
                    \mathbf{q}
                )
        \\
        &=
            \sum_{j}
                \sum_{i}
                    p_{ji}
                    \log\left(
                        \frac{
                            p_{ji}
                        }{
                            q_i
                        }
                    \right).
\end{align}
We can include the normalization of \(\mathbf{q}\) inside this function, so as we optimize it to solve for \(\mathbf{q}\), the resulting \(\mathbf{q}\) will automatically be normalized:
\begin{align}
    D_{\mathrm{cluster}}(
        \{\mathbf{p}_j\},
        \mathbf{q}
    )
        &=
            \sum_{j}
                \sum_{i}
                    p_{ji}
                    \log\left(
                        \frac{
                            p_{ji}
                        }{
                            q_i
                            /
                            \sum_k q_k
                        }
                    \right).
\end{align}
Taking the gradient of \(D_{\mathrm{cluster}}\) with respect to \(\mathbf{q}\) and setting all of its elements equal to zero results in the solution
\begin{align}
    \frac{
        q_i
    }{
        \sum_k q_k
    }
        &=
            q_i
        =
            \frac{
                \sum_j
                    p_{ji}
            }{
                \sum_{ji}
                    p_{ji}
            },
\end{align}
or in other words, that the cluster center of any set of normalized spectra is simply the element-wise mean of the normalized spectra.
To show this explicitly, we note that due to the normalization of the spectra,
\begin{align}
    \sum_{ji}
        p_{ji}
            &=
                \sum_{j=0}^{n-1}
                    \sum_i
                        p_{ji}
            =
                \sum_{j=0}^{n-1}
                    1
            =
                n,
\end{align}
and so the solution above is equivalent to the element-wise mean of the cluster:
\begin{align}
    q_i
        &\equiv
            \frac{1}{n}
            \sum_{j=0}^{n-1}
                p_{ji}.
\end{align}

\begin{figure}[t!]
\begin{center}
\includegraphics[width=0.99\columnwidth]{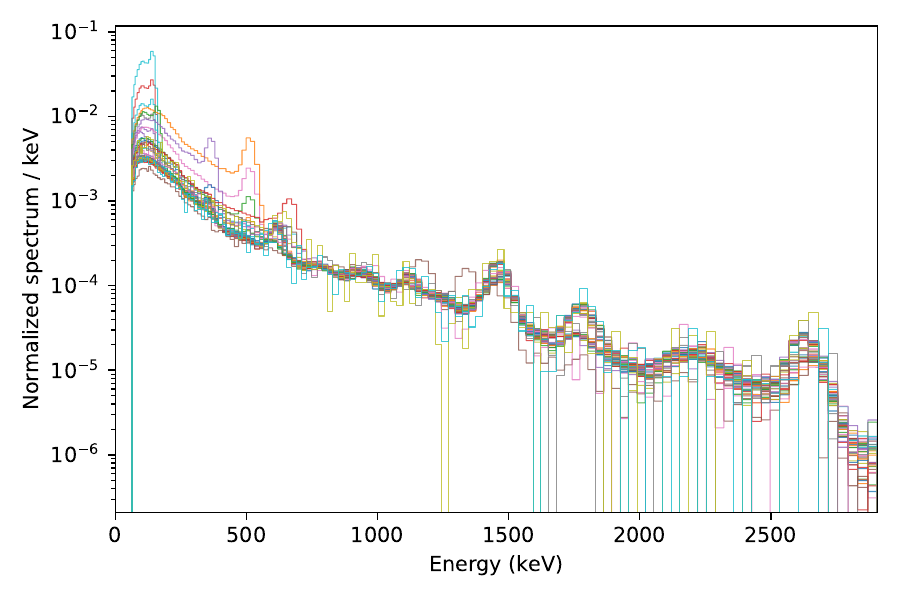}
\end{center}
\caption{Clusters of anomalous spectra found over 7~months of NOVArray detector 17178694 data using the KL divergence clustering method.
The cluster means are plotted so that their total counts above 1\,MeV are the same, making it easier to visually compare the cluster means with each other. \label{fig:kld-clusters}}
\end{figure}

As an example, the results of this procedure for anomalous spectra from NOVArray detector~17178694 for data taken between March 1, 2019 and September 30, 2019 are shown in~\Fref{fig:kld-clusters}.
Spectral triage was run on spectra with an integration time of 3\,s, and a total of 11,070 spectra were flagged as anomalous out of approximately \(6.16 \times 10^{6}\) spectra.
Various numbers of random clusters were initialized for K-means clustering, ranging from 10 to 100, and the results of each were reviewed qualitatively.
In the particular case shown, 40~random clusters were initialized.
The cluster means are plotted so that their total counts above 1\,MeV are the same, which makes them easy to visually compare to each other.
Visible are multiple clusters below \(\approx\)140\,keV for \isot{Tc}{99m}, multiple clusters for 511\,keV emission, a single cluster for \isot{Cs}{137} at 662\,keV, and a single cluster for \isot{Co}{60} at 1173 and 1332\,keV\@.

An advantage of the KL divergence clustering method is that it is simple to implement.
However, as was seen in this example, a disadvantage is that, due to the use of normalized spectra, many clusters could contain the same source type (e.g., \isot{Tc}{99m}), but at different strengths relative to background.
Additionally, if any background spectra are included in the set, due to, e.g., incorrectly alarming during rain events, those spectra will need to be included in clusters and can distort the process of finding anomaly clusters, especially for anomalies that only occur in a small number of spectra.

\subsection{Clustering using regularized NMF}
In order to cluster anomalies such that the same source at different strengths can belong to the same cluster, an alternative approach was explored.
This approach uses NMF with sparsity regularization to find a set of ``anomaly cluster components'' that, in addition to the static and rain background components, are able to fit the set of anomalous spectra.
The sparsity regularization will be applied only to the newly introduced components, to try to enforce the desired goal of expressing each anomalous spectrum as a linear combination of the two background components and only one of the anomalous components.

To implement the procedure, different numbers of anomaly cluster components (\( a \)) were randomly initialized and included in the NMF model \( \mathbf{W} \), in addition to the existing background model of \( b \) (usually 2) components.
Additional rows of \( \mathbf{H} \) were also added to match the number of columns of \( \mathbf{W} \).

To construct a regularization function, we considered the matrix \( \mathbf{G} \) where
\begin{align}
    G_{ij}
        &=
            \frac{
                S_{ij}
            }{
                \sqrt{
                    S_{ii}
                }
                \sqrt{
                    S_{jj}
                }
            },
\end{align}
where we have defined
\begin{align}
    S_{ij}
        &=
            \sum_{k=0}^{n-1}
                \left(
                    \frac{
                        H_{ik}
                    }{
                        \Delta t_k
                    }
                \right)
                \left(
                    \frac{
                        H_{jk}
                    }{
                        \Delta t_k
                    }
                \right),
\end{align}
making \( \mathbf{G} \) the matrix of all possible dot products of the rows of \( \mathbf{H} \), after each row has been converted to count rates and then normalized by its Euclidean norm.
Therefore, the diagonal of \( \mathbf{G} \) is always 1, and as a result of the non-negativity of \( \mathbf{H} \), \( G_{ij} = 1 \) if the shapes of the \(i\)-th and \(j\)-th rows of \( \mathbf{H} \) are the same (even if their overall magnitudes are different) and \( G_{ij} = 0 \) only if the weight assigned to component \( i \) is zero whenever the weight assigned to \( j \) is nonzero, and vice versa.

To make the anomaly cluster components sparse, in the sense that at most only one anomaly component should have a nonzero weight for each spectrum, we used the following regularization function to force the off-diagonal elements of \( \mathbf{G} \) to be close to zero:
\begin{align}
    f_3(\mathbf{H})
        &=
            \frac{n}{a (a-1)} \sum_{i,j=b}^{b+a-1} G_{ij},
\end{align}
i.e., only the sub-matrix of \( \mathbf{G} \) pertaining to the anomaly cluster components is summed over.
We have normalized the matrix sum by the total number of spectra \( n \) for the same reason as \( f_1 \) in~\fref{eq:f1}, which is to scale the regularized quantity by the number of spectra if it does not scale with \( n \) on its own.
We have also divided by the number of off-diagonal elements of the sub-matrix (\( a (a - 1) \), implicitly assuming \( a > 1 \)), since the sum should roughly scale with that number, so that weighting given for different \( a \) values is approximately the same.

We can express the gradient of \( f_3 \) with its non-negative parts
\begin{align}
    \left(
        \nabla_{\mathbf{H}}
        f_3
    \right)_{pq}^{+}
        &=
            2
            \left[ p \ge b \right]
            S_{pp}^{-1/2}
            \sum_{i=b}^{b+a-1}
                \frac{
                    H_{iq} / \Delta t_q^2
                }{
                    \sqrt{
                        S_{ii}
                    }
                }
        \\
    \left(
        \nabla_{\mathbf{H}}
        f_3
    \right)_{pq}^{-}
        &=
            2
            \left[ p \ge b \right]
            S_{pp}^{-3/2}
            \frac{
                H_{pq}
            }{
                \Delta t_q^2
            }
            \sum_{k=0}^{n-1}
                \frac{
                    H_{pk}
                }{
                    \Delta t_k
                }
                \sum_{i=b}^{b+a-1}
                    \frac{
                        H_{ik} / \Delta t_k
                    }{
                        \sqrt{
                            S_{ii}
                        }
                    },
\end{align}
where \( \left[ \cdot \right] \) is the Iverson bracket (1 if its argument is true, 0 otherwise).
We can then incorporate the gradients into the multiplicative update rules with a multiplication factor \( \alpha_3 \), analogously to what was previously shown in~\fref{eq:update_H_reg}.

As the multiplicative update rules are applied, since the first \( b \) components have been trained from background, they should not change, and thus the multiplicative update rules should not be applied to the first \( b \) columns of \( \mathbf{W} \).
It was found while tuning this method that it was important to also include the \(f_1\) regularization function, and additionally to apply \(f_1\) to the rain component's weights and not just the static component's.
Without these regularizations working in concert with \(f_3\), some anomaly cluster components would often resemble the static and rain components.
As before, the NMF model was fit by applying the multiplicative update rules until the same convergence criteria were met.
One difference from the previous use of regularizations is that since these regularizations were so strong, the unregularized multiplicative update rules (\fref{eq:update_W} and \fref{eq:update_H}) were used for the first 2,000 iterations before the regularizations were enabled, with the intention of allowing the randomized components to first fit to the statistics before being shaped by the regularization functions.

\begin{figure}[t!]
\begin{center}
\includegraphics[width=0.99\columnwidth]{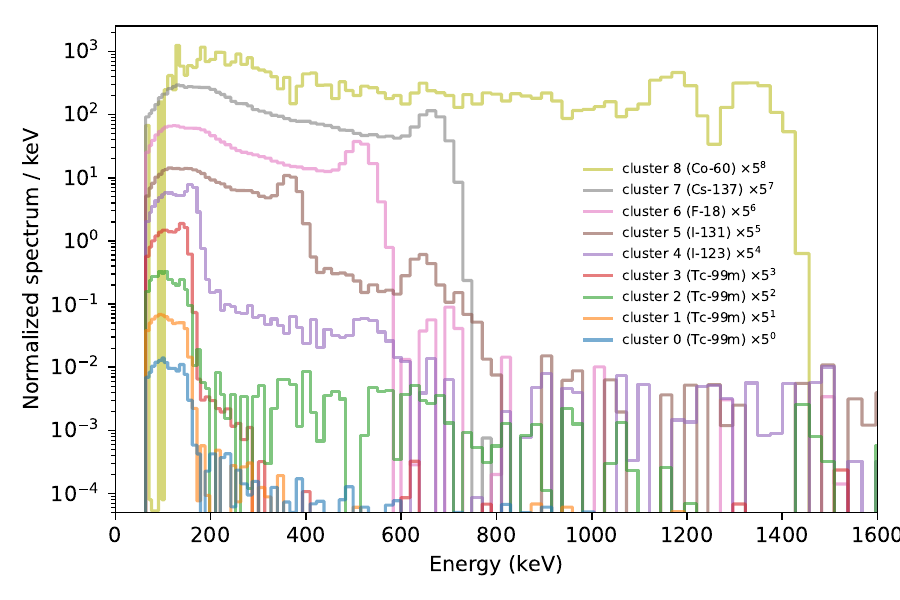}
\end{center}
\caption{Anomaly cluster components found using the NMF clustering method described in the text for seven months of data from NOVArray detector 17178694.
Tentative identifications are given using the minimum Jensen-Shannon divergence of each cluster to a library of simulated spectra.
\label{fig:nmf-clusters}}
\end{figure}

Using the same set of anomalies from NOVArray detector 17178694 as in~\Fref{sec:cluster-kldiv}, weighting factors of \( \alpha_1 = 10^{-2}\) and \(\alpha_3 = 10^{+3}\) were found to give interpretable and largely uncorrelated (off-diagonal elements of \( \mathbf{G} \) were all less than \( 4\times10^{-4} \)) anomaly clusters for different values of \(a\).
The best value of \(a\), in terms of giving the largest set of interpretable and mostly smooth anomaly spectra, was \(a = 9\), shown in~\Fref{fig:nmf-clusters}.
Monte Carlo-simulated source shapes were used to identify the anomaly cluster components, with the minimum Jensen-Shannon divergence being used to assign the best identification.
In this way, the anomalies were identified as \isot{Tc}{99m}, \isot{I}{123}, \isot{I}{131}, \isot{F}{18}, \isot{Cs}{137}, and \isot{Co}{60}.

When compared to the KL-divergence method of~\Fref{sec:cluster-kldiv}, now the spectral shapes of the anomalies are cleanly disentangled from background, and multiple source strengths can be summarized by a single anomaly cluster component.
The resulting spectra are also ready to be used without further processing in a template matching algorithm like NMF~\cite{bilton_nonnegative_2019}.
Also, since the previous algorithm uses K-means, it tends to find the largest clusters and ignore small clusters, but this method can be tuned to be sensitive to clusters that are present in only a small number of spectra --- in fact, by using the anomaly identification method of reference~\cite{bilton_nonnegative_2019}, we were able to estimate that only \(\approx\)17 spectra out of the entire dataset likely contained \isot{Co}{60}, which was the smallest number for any of the anomaly clusters.
And after many random initializations, we found that the \isot{Co}{60} component reliably appears for \(a \ge 9\), whereas in~\Fref{sec:cluster-kldiv} only very rarely did a \isot{Co}{60} cluster appear, and only for a large (\(>\) 30--40) number of clusters.

The nuclide \isot{Tc}{99m} was found in multiple anomaly clusters, each with a different shape, indicating that there is some variation among the shielding and/or scattering environments for that source.
Examining the cluster component shapes more closely, they appear to mostly differ in their ratios of the 140.5\,keV photopeak to the Compton continuum.
Appearing in multiple clusters may signal the need to develop multi-component models for \isot{Tc}{99m} for use in NMF or other template-matching algorithms.

\section{Implementation in the PANDAWN network}

The algorithms presented in the previous sections are at the core of the background model learning procedure developed for PANDA.
While developing a practical implementation, a number of issues and considerations arise when transitioning to an actual system in the field.
Additionally, computing resource limitations and real-time processing requirements can also constrain the implementation.
In this section, we briefly describe the major components of our implementation, alongside issues encountered during deployment, as well as possible solutions.
Suggested recommendations and technical implementation details will be the subject of a future publication.

Following are considerations about the background learning process:
\begin{itemize}
    \item Spectral triage is used to classify spectra into three categories: static background, rain, and anomalous.
    This scheme assumes that slow increases in gross count rate correlate with increased concentrations of radon progeny during rain events, while fast gross count rate increases correlate with sporadic source encounters.
    Accurate classification is required to compute distinct static background and rain-induced components for a simple one-component static model, and a two-component rain model, respectively.
    \item However, data acquired from the PANDA nodes has shown that radon enhancement does not only occur during rain events, but can also result from other phenomena such as temperature inversions.
    A method based on the detection of \isot{Ra}{226} anomalies was developed to flag this ``indirect'' radon enhancement contribution, and assure those spectra do not end up classified as static background.
    One notes that the ``direct'' (associated with rain events) and indirect radon enhancement contributions may result in distinct spectral components, in which case different models should be used to avoid any loss of sensitivity, at low energies in particular.
    \item Accurate classification is all the more important that spectra flagged as anomalous can be used to build data-based source templates for anomaly detection, using the clustering approach previously described.
\end{itemize}

Once a background model is learned, the question arises as to how to monitor the quality of this model over time, i.e., how to check that the background data does not deviate significantly from the model:
\begin{itemize}
    \item A possible adaptive approach is to make sure the model always fits the data by regularly retraining the model.
    Such an approach cannot easily be implemented on a system with limited computing resources, and real-time requirements.
    \item Instead, a method based on the AIC was developed to monitor the goodness-of-fit of the model over time.
    In this approach, retraining a model is only performed when a significant deviance in the AIC score of the model is observed.
    This ``train-as-required'' approach may however still put too much pressure on edge computing resources.
    For PANDA, an offload mechanism was developed to perform model training on the cloud.
    \item One also needs to consider which action to take when a retrained model does not perform better.
    In the PANDA implementation, the model gets retrained a second time, including an additional component to capture real physical changes in the background, or possible instrumentation effects.
    Beyond that point, experience has shown that human intervention may be necessary to investigate the system.
\end{itemize}

Finally, we review some considerations about the model selection process:
\begin{itemize}
    \item A multi-component model would eventually fit data acquired in different conditions reasonably well, avoiding the task of having to select specific models.
    This strategy has two disadvantages though: fitting non-necessary components can lead to a loss of sensitivity in source detection at low energy, and also requires the use of extra computation resources.
    \item The PANDA implementation relies on the assumption that the most significant change in background comes from radon enhancement in the data, whether this enhancement is direct (rain) or indirect (e.g. temperature inversions).
    Switching back and forth between these two classes of models is performed by constantly monitoring metrics correlating with the onset and offset of radon enhancement in the data.
\end{itemize}


\section{Discussion}

We have presented various physics-informed methods that can be used to understand and model the data from static spectroscopic detectors in real-world environments, thereby allowing them to dynamically adapt to their local radiological environment.
Special attention was paid to modeling the effects of rainfall using NMF, including information from rain sensors when available; clustering the spectral anomalies encountered by the system using two possible approaches; and relearning background models to accommodate detector performance and general background changes over time.
The models rely on some hyperparameters, which have been selected based on observation of desired performance based on the data evaluated.
It would be of significant value to this work and applications of this work to evaluate whether the hyperparameters are generally stable, or for example, whether environmental conditions such as different rain and snowfall patterns would significantly alter \(\alpha_1\) and \(\alpha_2\) in~\fref{eq:update_H_reg}.
Likewise, although we compared two clustering methods in~\fref{sec:anomaly-clustering}, we realize other clustering methods are worthy of consideration.

Some of the methods that have been presented have also been implemented in a real-time data processing framework and deployed as part of an experimental multi-sensor radiation detector network (the PANDAWN network) in Chicago.
The implementation includes additional methods to ascertain the ongoing validity of the spectral models in order to automatically account for shifts in the background due to, e.g., physical changes in the environment.
Methods such as those presented here are essential to enabling an ``active sensing'' concept for radiation detection in which detector systems are able to maintain performance by adapting to changing experimental conditions.

Data collection and analysis are ongoing, as well as validation and tuning of the adaptive background modeling approach in the deployed sensors.
The PANDA and DAWN projects are also currently performing several controlled experiments in which individual sensors nodes are exposed to gamma-ray sources under varying thermal, humidity, and precipitation scenarios.
The analysis of these data will serve to validate the performance and robustness of the various algorithms under well known conditions.

Related work on this project has included real-time, full-spectrum calibration, and simulation-based explorations of the detection performance of the PANDAWN network, anticipated to be available in the near future.


\section*{CRediT statement}

\textbf{M.S.B.}: conceptualization; formal analysis (lead); investigation (lead); methodology; software; writing -- original draft (lead); writing -- review \& editing.
\textbf{N.A.}: data curation; formal analysis; investigation; software; writing -- original draft (supporting); writing -- review \& editing.
\textbf{R.J.C.}: funding acquisition; project administration; supervision; writing -- review \& editing.
\textbf{D.H.}: methodology; software.
\textbf{T.H.Y.J}: supervision.
\textbf{V.N.}: data curation.
\textbf{B.J.Q.}: funding acquisition; supervision; writing -- review \& editing.
\textbf{M.S.}: data curation; software.
\textbf{R.S.}: supervision.
\textbf{Y.K.}: data curation.
\textbf{S.S.}: data curation.

\bibliographystyle{IEEEtran}
\bibliography{panda_modeling}

\end{document}

%% file: fig/matrix_diagram.tex
\begin{tikzpicture}[scale=0.8]
    \pgfmathsetmacro{\a}{0.3}
    \pgfmathsetmacro{\xleft}{0.0}
    \pgfmathsetmacro{\xright}{\xleft + 6.0}
    \pgfmathsetmacro{\xtop}{3.0}
    \pgfmathsetmacro{\xbottom}{\xtop - 3.0}

    \pgfmathsetmacro{\wleft}{\xright + 2.5}
    \pgfmathsetmacro{\wright}{\wleft + 1.2}
    \pgfmathsetmacro{\wtop}{\xtop}
    \pgfmathsetmacro{\wbottom}{\xbottom}

    \pgfmathsetmacro{\hleft}{\wright + 1.5}
    \pgfmathsetmacro{\hright}{\hleft + \xright - \xleft}
    \pgfmathsetmacro{\htop}{0.5 * (\wtop + \wbottom) + 0.5 * (\wright - \wleft)}
    \pgfmathsetmacro{\hbottom}{\htop - (\wright - \wleft)}

    \pgfmathsetmacro{\xcenterx}{0.5 * (\xleft + \xright)}
    \pgfmathsetmacro{\xcentery}{0.5 * (\xbottom + \xtop)}
    \pgfmathsetmacro{\wcenterx}{0.5 * (\wleft + \wright)}
    \pgfmathsetmacro{\wcentery}{0.5 * (\wbottom + \wtop)}
    \pgfmathsetmacro{\hcenterx}{0.5 * (\hleft + \hright)}
    \pgfmathsetmacro{\hcentery}{0.5 * (\hbottom + \htop)}

    \draw [very thick] (\xleft + \a, \xtop) -- (\xleft, \xtop) -- (\xleft, \xbottom) -- (\xleft + \a,\xbottom);
    \draw [very thick] (\xright - \a, \xtop) -- (\xright, \xtop) -- (\xright, \xbottom) -- (\xright - \a,\xbottom);

    \draw [very thick] (\wleft + \a, \wtop) -- (\wleft, \wtop) -- (\wleft, \wbottom) -- (\wleft + \a,\wbottom);
    \draw [very thick] (\wright - \a, \wtop) -- (\wright, \wtop) -- (\wright, \wbottom) -- (\wright - \a,\wbottom);

    \draw [very thick] (\hleft + \a, \htop) -- (\hleft, \htop) -- (\hleft, \hbottom) -- (\hleft + \a,\hbottom);
    \draw [very thick] (\hright - \a, \htop) -- (\hright, \htop) -- (\hright, \hbottom) -- (\hright - \a,\hbottom);

    \pgfmathsetmacro{\r}{0.20}  
    \pgfmathsetmacro{\e}{0.05}  
    \pgfmathsetmacro{\b}{0.08}  

    \pgfmathsetmacro{\nxx}{int(floor((\xright - \xleft - 2 * \b + 2 * \e) / (\r + 2 * \e)))}
    \pgfmathsetmacro{\nxy}{int(floor((\xtop - \xbottom - 2 * \b + 2 * \e) / (\r + 2 * \e)))}
    \pgfmathsetmacro{\nwx}{int(floor((\wright - \wleft - 2 * \b + 2 * \e) / (\r + 2 * \e)))}
    \pgfmathsetmacro{\nwy}{int(floor((\wtop - \wbottom - 2 * \b + 2 * \e) / (\r + 2 * \e)))}
    \pgfmathsetmacro{\nhx}{int(floor((\hright - \hleft - 2 * \b + 2 * \e) / (\r + 2 * \e)))}
    \pgfmathsetmacro{\nhy}{int(floor((\htop - \hbottom - 2 * \b + 2 * \e) / (\r + 2 * \e)))}

    \pgfmathsetmacro{\oxx}{0.5 * (\xright - \xleft - 2 * \b + 2 * \e - \nxx * (\r + 2 * \e))}
    \pgfmathsetmacro{\oxy}{0.5 * (\xtop - \xbottom - 2 * \b + 2 * \e - \nxy * (\r + 2 * \e))}
    \pgfmathsetmacro{\owx}{0.5 * (\wright - \wleft - 2 * \b + 2 * \e - \nwx * (\r + 2 * \e))}
    \pgfmathsetmacro{\owy}{0.5 * (\wtop - \wbottom - 2 * \b + 2 * \e - \nwy * (\r + 2 * \e))}
    \pgfmathsetmacro{\ohx}{0.5 * (\hright - \hleft - 2 * \b + 2 * \e - \nhx * (\r + 2 * \e))}
    \pgfmathsetmacro{\ohy}{0.5 * (\htop - \hbottom - 2 * \b + 2 * \e - \nhy * (\r + 2 * \e))}

    \foreach \x in {1,...,\nxx}
        \foreach \y in {1,...,\nxy}
            \fill [black!10] ({\xleft + \oxx + \b + (\x - 1) * \r + 2 * (\x - 1) * \e}, {\xbottom + \oxy + \b + (\y - 1) * \r + 2 * (\y - 1) * \e}) rectangle ++(\r, \r);

    \foreach \x in {1,...,\nwx}
        \foreach \y in {1,...,\nwy}
            \fill [black!10] ({\wleft + \owx + \b + (\x - 1) * \r + 2 * (\x - 1) * \e}, {\wbottom + \owy + \b + (\y - 1) * \r + 2 * (\y - 1) * \e}) rectangle ++(\r, \r);

    \foreach \x in {1,...,\nhx}
        \foreach \y in {1,...,\nhy}
            \fill [black!10] ({\hleft + \ohx + \b + (\x - 1) * \r + 2 * (\x - 1) * \e}, {\hbottom + \ohy + \b + (\y - 1) * \r + 2 * (\y - 1) * \e}) rectangle ++(\r, \r);

    \node at (\xcenterx, \xcentery) {\LARGE $\mathbf{X}$};
    \node at ({0.5 * (\xright + \wleft)}, \xcentery) {\huge $\approx$};
    \node at (\wcenterx, \wcentery) {\LARGE $\mathbf{W}$};
    \node at ({0.5 * (\wright + \hleft)}, \xcentery) {\huge $\cdot$};
    \node at (\hcenterx, \hcentery) {\LARGE $\mathbf{H}$};

    \pgfmathsetmacro{\c}{0.15}

    \draw [decorate, decoration={brace, mirror, amplitude=10pt}]
    (\xleft + \c, \xbottom - \c) -- (\xright - \c, \xbottom - \c) node [black, midway, yshift=-0.6cm, font=\footnotesize\linespread{0.9}\selectfont] {columns are spectra ($\mathbf{x}_i$)};

    \draw [decorate, decoration={brace, mirror, amplitude=10pt}]
    (\wleft + \c, \wbottom - \c) -- (\wright - \c, \wbottom - \c) node [black, midway, yshift=-0.6cm, align=center, text width=4.0cm, font=\footnotesize\linespread{0.9}\selectfont] {columns are components ($\mathbf{w}_j$)};

    \draw [decorate, decoration={brace, mirror, amplitude=10pt}]
    (\hleft + \c, \hbottom - \c) -- (\hright - \c, \hbottom - \c) node [black, midway, yshift=-0.6cm, text width=4.0cm, font=\footnotesize\linespread{0.9}\selectfont] {columns are \(\mathbf{h}_i\) and rows are \(\mathbf{h}_j^{\top}\)};

    \draw [-{Latex}] (\xleft - \a, \xtop - \oxy - \b) -- (\xleft - \a, \xbottom + \b) node [midway, left, xshift=-0.3cm, text centered, anchor=center, rotate=90, font=\footnotesize\linespread{0.9}\selectfont] {spectral bins (length $m$)};

    \draw [-{Latex}] (\xleft + \oxx + \b, \xtop + \a) -- (\xright - \b, \xtop + \a) node [midway, above, yshift=0.1cm, text centered, font=\footnotesize\linespread{0.9}\selectfont] {measurement (index $i$, length $n$)};

    \draw [-{Latex}] (\wleft + \owx + \b, \wtop + \a) -- (\wright - \b, \wtop + \a) node [midway, above, yshift=0.1cm, text centered, text width=4.0cm, font=\footnotesize\linespread{0.9}\selectfont] {component\\ (index $j$, length $d$)};

    \draw [-{Latex}] (\hleft + \ohx + \b, \htop + \a) -- (\hright - \b, \htop + \a) node [midway, above, yshift=0.1cm, text centered, font=\footnotesize\linespread{0.9}\selectfont] {measurement (index $i$, length $n$)};
\end{tikzpicture}